\documentstyle[preprint,aps,floats,epsf]{revtex}
\begin{document}
\tightenlines
\preprint{\font\fortssbx=cmssbx10 scaled \magstep2
\hbox to \hsize{
\hskip.5in \raise.1in\hbox{\fortssbx University of Wisconsin - Madison}
\hfill$\vcenter{\hbox{\bf MADPH-95-892}
                \hbox{June 1995}}$}}

\title{\vspace*{.5in}
Probing Trilinear Gauge Boson Couplings at Colliders
\footnote{Talk given at the {\it International Symposium on Vector Boson
Self-Interactions}, Feb. 1--3, 1995, UCLA, Los Angeles}
}

\author{Dieter Zeppenfeld}
\address{Department of Physics, University of Wisconsin, Madison, WI 53706}

\maketitle

\vspace{1in}

\begin{abstract}
A direct measurement of the trilinear $WW\gamma$ and $WWZ$ couplings is
possible in the pair production of electroweak bosons at $e^+e^-$ and hadron
colliders. This talk addresses some of the theoretical issues: the
parameterization of ``anomalous couplings'' in terms of form factors and
effective Lagrangians, the complementary information which can be obtained
in $e^+e^-$ vs. hadron collider experiments, and a novel way to implement
finite $W$-width effects in a gauge invariant manner.
\end{abstract}

\thispagestyle{empty}
\newpage

\section*{1. Introduction}

Over the past twenty years a wealth of high precision electroweak data
has beautifully confirmed the SM predictions for the couplings of fermions
to the electroweak gauge bosons. Measurements of the $f\bar f V$ couplings
at LEP and the SLC generally agree with the SM at the 0.1--1\%
level\cite{schaile} and universality of the lepton couplings has been tested
at a similar level. This agreement provides strong evidence that the gauge
theory description of electroweak interactions is indeed correct.
In spite of these successes the most direct consequence of the underlying
$SU(2)$ gauge symmetry, the nonabelian couplings of photons, $Z$'s and $W$'s,
remain to be tested with meaningful precision.

Pair production of electroweak bosons ($W^+W^-$ production at $e^+e^-$
colliders, $W\gamma$, $WZ$ and $W^+W^-$ production at hadron colliders) are
the prime processes to directly measure the $WWV,\; V=\gamma,Z$ couplings.
With high enough precision one may hope to be sensitive to new physics in
the bosonic sector. However, one likely will need a lepton collider in the
TeV range to reach the required sensitivity~\cite{dpf,hernandez}.
For machines such as the Tevatron or LEP II the foremost task will be to
confirm the SM predictions for the $WWV$ couplings and to quantify this
agreement. For both purposes, discovery of new physics and SM tests, one can
introduce a $WWV$ vertex with generalized coupling parameters and then
experimentally constrain their deviations from the SM predictions. This is
analogous to the introduction of axial and vector couplings $g_A$ and $g_V$
for the $f\bar f V$ vertex. In Section~2 I will discuss parameterizations of
the $WWV$ vertices, both in terms of form-factors and effective
Lagrangians. Ways to extract these couplings from data on weak boson pair
production will be discussed in Section~3, with special emphasis on the
complementarity of hadron and $e^+e^-$ colliders. Many of these questions
are considered in greater detail in other contributions to these Proceedings
so the discussion here will be limited to some of the more fundamental
questions.

Experimentally one only observes the decay products of $W$'s and $Z$'s and
finite width effects must be taken into account, in particular when working
close to threshold, such as at the Tevatron or at LEP II. Implementing finite
width effects while maintaining gauge invariance becomes a nontrivial task
when nonabelian couplings are present. These
questions will be discussed in Section~4 for the example of $W\gamma$
production in $p\bar p$ collisions. It is shown how inclusion of fermion
triangle graphs together with resummation of vacuum polarization
contributions (which are the basis for the $W$ Breit Wigner propagator) lead
to a gauge invariant result. At the same time this example of SM radiative
corrections will serve to illustrate some of the points made in the previous
Sections.

\section*{2. Anomalous Couplings and Form-Factors}

Because of the rapid decay of $W$'s and $Z$'s, weak boson pair production
is seen experimentally as the process
\begin{equation}
f_1\bar f_2 \to V_1V_2 \to f_3\bar f_4\; f_5\bar f_6\; ,
\end{equation}
with both final state fermion-antifermion pairs in a $J=1$ angular momentum
state. We are thus interested in deviations $\Delta {\cal M}$ from SM
six-fermion amplitudes ${\cal M}_{6f}$ in particular partial waves.
Apart from anomalies in the
three gauge boson couplings such deviations may also arise from new physics
in the gauge boson--fermion interactions or from non-standard behaviour
of the gauge boson propagators. The latter two, however, are already
tested, at the 1\% level or slightly better, in four-fermion processes like
$e^+e^- \to f\bar f$ and we thus assume SM behaviour for both.
\footnote{This implies that once three boson couplings
are tested beyond $10^{-2}$ accuracy the assumption of SM behaviour
of propagators and fermion vertices should be revisited.}
We are left with deviations $\Delta {\cal M}$ which occur
due to the Three Gauge Vertex (TGV) and therefore appear in the overall
$J=1$ partial wave. Denoting the decay currents by {\it e.g.}
\begin{equation}
J^{(34)}_{V_1\alpha}(q) = \bar f_3\gamma^{\alpha'}
(g_V^{f_3f_4V_1}+g_A^{f_3f_4V_1}\gamma_5)
f_4 \; D^{V_1}_{\alpha'\alpha}(q) \; ,
\end{equation}
where the gauge boson propagator $D^{V}$ has been included in the definition
of the current, we may write the deviation $\Delta {\cal M}$ as
\begin{equation}
\Delta {\cal M} = \sum J^{(12)}_{V\mu}(P)\;
J^{(34)}_{V_1\alpha}(q) \; J^{(56)}_{V_2\beta}(\bar q)\;
g_{VV_1V_2}\; \Delta \Gamma^{\mu\alpha\beta}_{VV_1V_2}(P,q,\bar q)\; .
\end{equation}
Here $g_{WW\gamma}=-e$, $g_{WWZ}=-e\,{\rm tan}\theta_W$ and the sum
indicates that for neutral currents we need to add photon and $Z$ exchange.

By convention we include the SM tree level vertex in the definition of the
vertex function $\Gamma^{\mu\alpha\beta}_{VV_1V_2}(P,q,\bar q)$.
The momentum assignment for the vertex function is depicted in
Fig.~1. 
In the limit of massless external fermions the
currents $J^{(ij)}_{V\mu}$ are conserved, {\it i.e.} terms like
$P^\mu J^{(12)}_{V\mu}(P)$ can be neglected. As a result
the most general tensor structure of the vertex function can be written in
terms of seven form factors $f_i(P^2,q^2,\bar q^2)$ \cite{GG,HPZH}
\begin{eqnarray}
\Gamma^{\mu\alpha\beta}(P,q,\bar q)= &f_1& (q-\bar q)^\mu g^{\alpha\beta}
-{f_2\over m_W^2}(q-\bar q)^\mu P^\alpha P^\beta
+f_3(P^\alpha g^{\mu\beta} - P^\beta g^{\mu\alpha})  \nonumber \\
+&if_4&(P^\alpha g^{\mu\beta} + P^\beta g^{\mu\alpha})
+if_5 \varepsilon^{\mu\alpha\beta\rho}(q-\bar q)_\rho \nonumber \\
-&f_6& \varepsilon^{\mu\alpha\beta\rho} P_\rho
-{f_7\over m_W^2}(q-\bar q)^\mu \varepsilon^{\alpha\beta\rho\sigma}
P_\rho (q-\bar q)_\sigma \; .
\label{Gamdef}
\end{eqnarray}
This decomposition is completely general and applicable at all energies.
Discrete symmetries of the underlying dynamics imply constraints among them.
Parity conservation leads to $f_5 = f_6 = f_7 = 0$. Charge conjugation
invariance relates the form factors $f_1$, $f_2$, $f_3$, and $f_4$.

\begin{figure}[h]
\epsfxsize=3.2in
\epsfysize=0.95in
\label{figGammaVVV}
\begin{center}
\hspace*{0in}
\epsffile{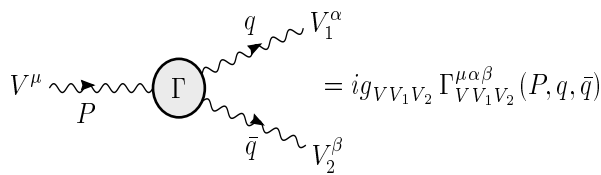}
\vspace*{0.1in}
\caption{Feynman rule for the general $V\to V_1V_2$ vertex.}
\vspace*{-0.15in}
\end{center}
\end{figure}

While the decomposition into form factors is general, convenient
parameterizations of their low-energy behavior are provided by the
effective Lagrangian approach~\cite{wudka}. Many different forms have
been used in the literature. Here it suffices to use the phenomenological
Lagrangian of Ref.~\cite{HPZH} as an example. Keeping $C$ and $P$ conserving
terms only, the $WWV$ vertex ($V=Z,\gamma$) is given in terms of three
parameters, $g_1^V$, $\kappa_V$ and $\lambda_V$,
\begin{eqnarray}\label{LeffWWV}
{\cal L}_{eff}^{WWV} & = & i\, g_{WWV} \Bigl(\; g_1^V (
W^+_{\mu\nu}W^{-\,\mu}-W^{+\, \mu}W^-_{\mu\nu})V^{\nu} +
\kappa_V\,  W^{+}_{\mu}W^-_{\nu}V^{\mu\nu}
 \nonumber \\
&& {\phantom {i\, g_{WWV} }}
+{\lambda_V\over m_W^2} W_{\mu}^{+\,\nu}W_{\nu}^{-\,\rho}
{V_{\rho}}^{\mu}\Bigr)\;,
\end{eqnarray}
where {\it e.g.} $V^{\mu\nu}= \partial^\mu V^\nu - \partial^\nu V^\mu$ is the
$\gamma$ or $Z$ field strength tensor. Within the SM, the couplings
are given by $g_1^Z = g_1^\gamma = \kappa_Z = \kappa_\gamma = 1$, and
$\lambda_Z  = \lambda_{\gamma}  = 0$. The effective Lagrangian of
Eq.~\ref{LeffWWV} provides us with the lowest order terms in an expansion
of the form factors $f_i$ in powers of the Lorentz invariants $P^2$, $q^2$
and $\bar q^2$. For (on- or off-shell) $W^+W^-$ production they are given by
\begin{eqnarray}
f_1^V(P^2,q^2,\bar q^2)\; & \approx & \;
g_1^V + \, \lambda_V  {P^2 \over 2 m_W^2}\; , \label{relftog1} \\
f_2^V(P^2,q^2,\bar q^2) \; & \approx & \;
\lambda_V \; ,\label{relftog2} \\
f_3^V(P^2,q^2,\bar q^2)\; & \approx & \;
g_1^V +\, \kappa_V + \, \lambda_V {q^2+\bar q^2 \over 2 m_W^2}\; ,
\label{relftog3} \\
f_4^V(P^2,q^2,\bar q^2)\; & \approx & \;
-i\lambda_V {q^2-\bar q^2 \over 2 m_W^2}\; .\label{relftog4}
\end{eqnarray}

The notation developed up to here is getting cumbersome when comparing
crossing related processes. The tensor decomposition of Eq.~\ref{Gamdef}
treats incoming and outgoing vector bosons differently. As a result the
form factors $f_i(P^2,q^2,\bar q^2)$ are process dependent: they mix under
crossing. It is more convenient to define form factors
$g_1^V(P^2,q^2,\bar q^2)$, $\kappa_V(P^2,q^2,\bar q^2)$ and
$\lambda_V(P^2,q^2,\bar q^2)$ such that the relations of
Eqs.~\ref{relftog1}--\ref{relftog3}
become exact for $V(P)\to W^-(q)W^+(\bar q)$. This approach has the
advantage that the Feynman rules derived from the effective Lagrangian
of Eq.~\ref{LeffWWV} can be used directly to calculate the full form
factor dependence for any process involving $WWV$ vertices. At the same time
the relations between form factors for crossing related processes become
manifest. The only disadvantage is that the coupling constants appearing
in the effective Lagrangian might be confused with the full form factors,
while in reality they just represent the low energy limits of these form
factors.

The functional behaviour of the form factors depends on the
details of the underlying new physics. Effective Lagrangian
techniques~\cite{wudka} are of little help here because the low energy
expansion which leads to the effective Lagrangians exactly breaks down where
the form factor effects become important. So in practice one will have to
make ad hoc assumptions. One possibility is to assume a behaviour similar to
nucleon form factors, with constraints derived from unitarity
considerations~\cite{BZunitarity}. Such constraints do become important
at hadron colliders. More generally, they must be included when one searches
for very large enhancements of vector boson pair production cross sections.

\section*{3. Vector Boson Pair Production}

Deviations of the TGV's from their SM, tree level form are most directly
observed in vector boson pair production. Candidate processes are $W\gamma$,
$WZ$ and $W^+W^-$ production at hadron colliders (namely the Tevatron and,
eventually, the LHC) and $e^+e^-\to W^+W^-$ at LEP II or a NLC. Since
experimental strategies have been discussed at great depth by other speakers
at this symposium~\cite{VVprod}, I will concentrate here on some of the more
basic effects of anomalous TGV's on vector boson pair production.

\subsection*{3.1 $W^+W^-$ Production in $e^+e^-$ Collisions}

To lowest order, the production of $W$ pairs in $e^+e^-$ collisions proceeds
via the Feynman graphs of Fig.~2. 
It is instructive to consider the
individual contributions of $s$-channel photon and $Z$  exchange and of
$t$-channel neutrino exchange to the various helicity amplitudes~\cite{HPZH},
\begin{figure}[h]
\epsfxsize=4.5in
\epsfysize=1.0in
\label{figww}
\begin{center}
\hspace*{0in}
\epsffile{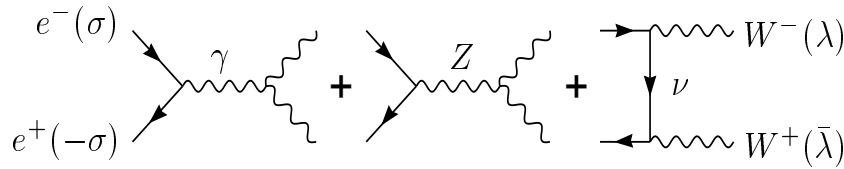}
\vspace*{0.1in}
\caption{Feynman graphs for the process $e^+e^-\to W^+W^-$.}
\vspace*{-0.1in}
\end{center}
\end{figure}
\begin{equation}
{\cal M}(\sigma,\lambda,\bar\lambda) = {\cal M} =
{\cal M}_\gamma + {\cal M}_Z + {\cal M}_\nu \; .
\end{equation}
Here the $e^-$ and $e^+$ helicities are given by $\sigma/2$ and $-\sigma/2$,
and $\lambda$ and $\bar \lambda$ denote the $W^-$ and $W^+$ helicities.
Following Ref.~\cite{HPZH} let us define reduced amplitudes $\tilde{\cal M}$
by splitting off the leading angular dependence in terms of the $d$-functions
$d^{J_0}$ where $J_0= 1,2$ denotes the lowest angular momentum
contributing to a given helicity combination,
\begin{equation}
{\cal M}(\sigma,\lambda,\bar\lambda;\theta) = \sqrt{2} e^2
\tilde{\cal M}_{\sigma,\lambda,\bar\lambda}(\theta)\;
d^{J_0}_{\sigma,\lambda-\bar\lambda}(\theta)\; .
\label{Mdf}
\end{equation}
$s$-channel photon and $Z$ exchange is only possible for
$|\lambda-\bar\lambda| = 0,1$. The corresponding reduced amplitudes can
be written as
\begin{eqnarray}
\tilde{\cal M}_\gamma & = &
-\beta A^\gamma_{\lambda\bar\lambda} \; , \nonumber \\
\tilde{\cal M}_Z & = &
+\beta A^Z_{\lambda\bar\lambda} \left[
1-\delta_{\sigma,-1}{1\over 2\, {\rm sin}^2 \theta_W }
\right] {s \over s-m_Z^2}\; , \nonumber \\
\tilde{\cal M}_\nu & = &
+\delta_{\sigma,-1}{1\over 2\beta\, {\rm sin}^2 \theta_W }
\left[B_{\lambda\bar\lambda} -
{1\over 1+\beta^2-2\beta {\rm cos} \theta} C_{\lambda\bar\lambda} \right]
\; . \label{Mreduced}
\end{eqnarray}
Here $s$ denotes the $e^+e^-$ center of mass energy and
$\beta = \sqrt{1-4m_W^2/s}$ is the $W^\pm$ velocity. The subamplitudes
$A^V$, $B$ and $C$ are given in Table~\ref{tableamp}.
\begin{table}[t]
\caption{
Subamplitudes for $J_0=1$ helicity combinations of the process
$e^-e^+\to W^-W^+$, as defined in Eq.~\protect\ref{Mreduced}. $\beta$
denotes the $W$ velocity and $\gamma = \protect\sqrt{s}/2m_W$. The
relations between the form factors $f_i$ and $g_1$, $\kappa$, and
$\lambda$ are given in
Eqs.~\protect\ref{relftog1}--\protect\ref{relftog3},
with $q^2=\bar q^2 = m_W^2$.
}
\label{tableamp}
\begin{tabular}{lddd}
\multicolumn{1}{c}{$\lambda\bar\lambda$}& $A^V_{\lambda\bar\lambda}$ &
   $B_{\lambda\bar\lambda}$ & $C_{\lambda\bar\lambda}$\\
\tableline
$++$ & $g_1^V +2\gamma^2\lambda_V +{i\over\beta}f_6^V +4i\gamma^2\beta f_7^V$
& 1 & $1/ \gamma^2$ \\
$--$ & $g_1^V +2\gamma^2\lambda_V -{i\over\beta}f_6^V -4i\gamma^2\beta f_7^V$
& 1 & ${1/ \gamma^2}$ \\
$+0$ & $\gamma (f_3^V -if_4^V+\beta f_5^V + {i\over \beta}f_6^V )$
& $2\gamma$ & $2(1+\beta)/\gamma$ \\
$0-$ & $\gamma (f_3^V +if_4^V+\beta f_5^V - {i\over \beta}f_6^V )$
& $2\gamma$ & $2(1+\beta)/\gamma$ \\
$0+$ & $\gamma (f_3^V +if_4^V-\beta f_5^V + {i\over \beta}f_6^V )$
& $2\gamma$ & $2(1-\beta)/\gamma$ \\
$-0$ & $\gamma (f_3^V -if_4^V-\beta f_5^V - {i\over \beta}f_6^V )$
& $2\gamma$ & $2(1-\beta)/\gamma$ \\
$00$ & $g_1^V +2\gamma^2\kappa_V$
& $2\gamma^2$ & ${2/ \gamma^2}$ \\
\end{tabular}
\end{table}

One of the most striking features of the SM are the gauge theory cancellations
between $\gamma$, $Z$ and neutrino exchange graphs. Within the SM $g_1 = \kappa
=1,\; \lambda = 0$ (or $f_1 = 1$, $f_2=0$, $f_3=2$) for both the photon and
the $Z$-exchange graphs. As a result $A^\gamma_{\lambda\bar\lambda}=
A^Z_{\lambda\bar\lambda}$ and the $\beta A^V$ terms in Eq.~\ref{Mreduced}
cancel, except for the difference between photon and $Z$ propagators.
Similarly, the $B_{\lambda\bar\lambda}$ term in $\tilde{\cal M}_\nu$ and the
$\delta_{\sigma,-1}$ term in $\tilde{\cal M}_Z$ cancel in the high energy
limit for all helicity combinations. While the contributions from individual
Feynman graphs grow with energy for longitudinally polarized $W$'s, this
unacceptable high energy behavior is avoided in the full amplitude due to the
cancellations which can be traced to the gauge theory relations between
fermion--gauge boson vertices and the TGV's.

At asymptotically large energies any deviations of $f_3$, ... $f_6$ from their
SM values would lead to a growth of at least some of the helicity amplitudes
$\tilde{\cal M}_{0\pm}$
or $\tilde{\cal M}_{\pm0}$ with energy and hence violate partial wave
unitarity. Similarly, non-standard values of $f_7$, $\lambda$ or
$\kappa$ in the $s\to \infty$ limit would lead to an unacceptable
growth in some of the remaining three helicity amplitudes. Thus, in this limit,
partial wave unitarity excludes anomalous TGV's~\cite{joglekar}; any
deviation from the SM must be described by an energy-dependent form factor
which approaches its gauge theory value as $s\to \infty$.

Table~\ref{tableamp} shows that only seven $W^-W^+$ helicity combinations
contribute to the $J=1$ channel and the form factors $f_i$ enter in as
many different combinations. This explains why exactly
seven form factors or coupling constants are needed to parameterize the
most general $WWV$ vertex. Since we have both $WWZ$ and $WW\gamma$ couplings
at our disposal, the most general $J=1$ amplitudes
${\cal M}_L={\cal M}(\sigma=-1,\lambda,\bar\lambda)$ and
${\cal M}_R={\cal M}(\sigma=+1,\lambda,\bar\lambda)$ for both
left- and right-handed incoming electrons can be parameterized. Turning the
argument around one concludes that
all 14 helicity amplitudes need to be measured independently for a complete
determination of all the form factors $f_i^\gamma(s)$ and $f_i^Z(s)$, at any
value of the center of mass energy, $\sqrt s$.

Formidable as this goal may be it can be approached to a remarkable degree
by performing a partial wave analysis, in particular of the semileptonic
process $e^-e^+\to W^-W^+\to \ell^\pm\nu q\bar q'$. The charge of the lepton
allows to identify the two $W$ charges and hence the production angle $\theta$.
{}From Eq.~\ref{Mdf} one finds that the $J=1$ amplitudes lead to the angular
distribution
\begin{eqnarray}
{d\sigma\over d {\rm cos}\theta} &\sim & {{\rm sin}^2\theta\over 2}
\left( |\tilde {\cal M}_{\sigma,++}|^2+|\tilde {\cal M}_{\sigma,--}|^2
+|\tilde {\cal M}_{\sigma,00}|^2 \right)  \nonumber \\
&& + {(1+\sigma\, {\rm cos}\, \theta)^2\over 4}
\left( |\tilde {\cal M}_{\sigma,+0}|^2+|\tilde {\cal M}_{\sigma,0-}|^2 \right)
\nonumber \\
&& + {(1-\sigma\, {\rm cos}\, \theta)^2\over 4}
\left( |\tilde {\cal M}_{\sigma,0+}|^2+|\tilde {\cal M}_{\sigma,-0}|^2 \right)
\; .
\end{eqnarray}
Hence the amplitudes with different values of $|\lambda-\bar\lambda|$ can be
separated, even though in practice one must take into account the additional
$\theta$-dependence of the known neutrino exchange graphs, as is evident from
Fig.~3. 

\begin{figure}[h]
\epsfxsize=5in
\label{sighel}
\begin{center}
\hspace*{0in}
\epsffile{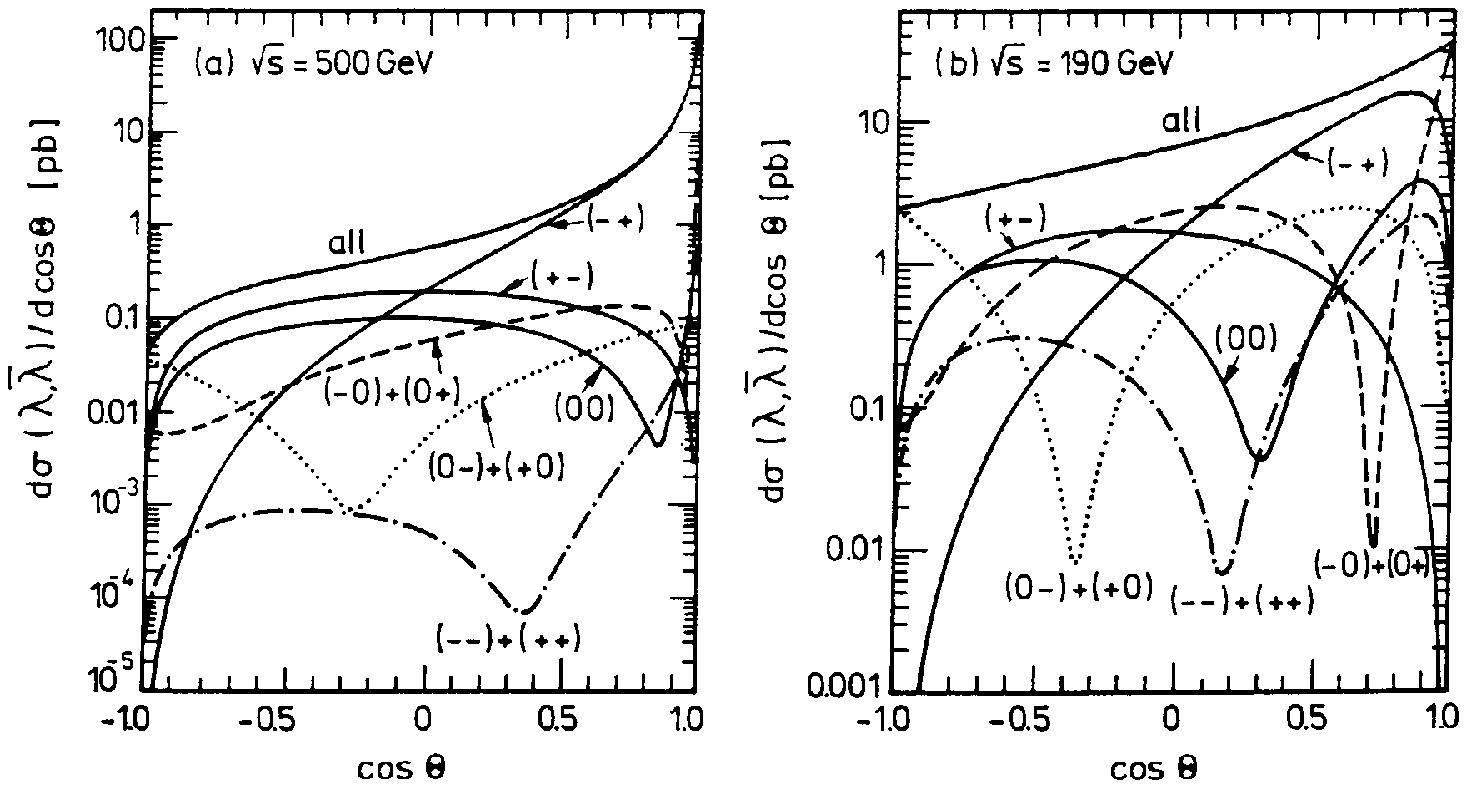}
\vspace*{0.1in}
\caption{Angular distributions $d\sigma/d{\rm cos}\theta$ for fixed $W^-W^+$
helicities $(\lambda\bar\lambda)$ in $e^-e^+$ collisions at a)
$\protect\sqrt{s} = 500$~GeV and b) $\protect\sqrt{s} = 190$~GeV.
{}From Ref.~\protect\cite{HPZH}. }
\vspace*{-0.15in}
\end{center}
\end{figure}

Due to the $V-A$ structure of the $W$--fermion vertices the decay angular
distributions of the $W$'s are excellent polarization analyzers. Consider
for example the polar angle $\theta_-$ of the charged lepton $\ell^-$ in
the $W^-$ rest frame with respect to the $W^-$ direction in the lab frame.
Its distribution is proportional to $(1-\lambda{\rm cos}\theta_-)^2$
for transversely polarized $W^-$ and proportional to ${\rm sin}^2\theta_-$
for longitudinally polarized $W^-$. Combined with the information contained
in the production angle distribution, the individual helicity amplitudes
can be isolated, at least when polarized electron beams are available.

In practice, statistical errors will limit the accuracy with which such an
analysis can be carried out. The best sensitivity to anomalous contributions
is achieved when interference with large SM amplitudes can be exploited.
Unfortunately, the dominant SM amplitudes are the $J_0=2$ amplitudes
${\cal M}_{+-}$ and ${\cal M}_{-+}$ which are purely due to $t$-channel
neutrino exchange (see Fig.~3). 
At asymptotically large energies
the only surviving $J=1$ helicity amplitude is ${\cal M}_{00}$ and even its
contribution is small numerically. Polar angle distributions alone yield
relatively low sensitivity to anomalous TGV's.

The way out is to measure azimuthal angle distributions and azimuthal angle
correlations which exploit the interference of the various $J=1$ amplitudes
with the $t$-channel neutrino exchange graph. In order to independently
measure the various form-factors it is necessary to measure the full
five-fold angular distributions
\begin{equation}
{d^5\sigma \over d{\rm cos}\theta\; d{\rm cos}\theta_+\; d\phi_+\;
d{\rm cos}\theta_-\; d\phi_-}\; ,
\end{equation}
or more precisely the projection of this five-fold angular distribution on the
triple $J=1$ partial wave.

The sensitivity of such an analysis has been
investigated for both LEP II and NLC energies~\cite{schild,barklow}.
One finds, for example, that $\Delta\kappa$ should be measurable with an
accuracy of $\approx 5\cdot 10^{-4}$ at a $1.5$~TeV NLC~\cite{barklow}.
Does this mean that
electroweak radiative corrections will be probed by measuring $W^+W^-$
production at a NLC? In spite of the small value of $\Delta\kappa$ this is
not necessarily the case. According to Table~\ref{tableamp}, an anomalous
value of $\Delta\kappa$ has its largest effect on $W_LW_L$ production. With
$\gamma=\sqrt{s}/2m_W\approx 10$ and taking the gauge theory cancellations
into account, the effect of $\Delta\kappa=5\cdot 10^{-4}$ on ${\cal M}_{00}$
is proportional to
\begin{equation}
\tilde{\cal M}_{00}\sim {m_Z^2\over 2 m_W^2}+2\gamma^2\Delta\kappa
=0.65 + 2\cdot 10^2\cdot 5\cdot 10^{-4} = 0.65\cdot 1.15\; ,
\end{equation}
$i.e.$ the anomalous TGV corresponds to a change of the $W_LW_L$ amplitude of
15\%, which probably is more than should be expected from electroweak
radiative corrections.

\subsection*{3.2 Weak Boson Pair Production at Hadron Colliders}

There are substantial differences in the study of TGV's at $e^+e^-$ vs.
hadron colliders. At LEP II or a NLC a detailed study of individual
helicity amplitudes is possible  and hence the individual form factors
can be separated at any center of mass energy $\sqrt{s}$. In the clean
environment of these machines errors are largely dominated by statistics
and weak boson pair production cross sections can hence be measured with
errors in the few percent range, i.e. the search for anomalous TGV's
corresponds to the search of ${\cal O}(10^{-2})$ deviations of the
production cross sections from the SM predictions.

Hadron colliders like the Tevatron or the LHC allow to study all pair
production processes: $W^+W^-$, $W^\pm\gamma$, and $W^\pm Z$ production.
Via the last two processes one can thus independently measure $WW\gamma$
and $WWZ$ vertices. At the same time larger center of mass energies are
available at the hadron machines compared to their $e^+e^-$ contemporaries
and hence the form factors are explored at higher energy scales. In turn
this implies large enhancement factors ($\gamma$ or $\gamma^2$) for the
anomalous contributions. The signals we are searching for appear in the $J=1$
partial wave $i.e.$ for large production angles of the final state
electroweak bosons and they are enhanced at large c.m. energies. Both features
move observable effects to large transverse momenta of the produced vector
bosons or their decay products. This effect is demonstrated in
Fig.~4 
where expected transverse momentum distributions in
$p\bar p\to W^+\gamma$ and $p\bar p\to W^+Z$ production at the Tevatron are
shown for several choices of anomalous couplings and dipole form factors
$\Delta f_i(s) = \Delta f_i^0/(1+s/1{\rm TeV}^2)^2$.

\begin{figure}
\label{figTeVWV}
\input rotate
\begin{center}
\hspace*{0in}
\setbox1\vbox{\epsfysize=5in\epsffile{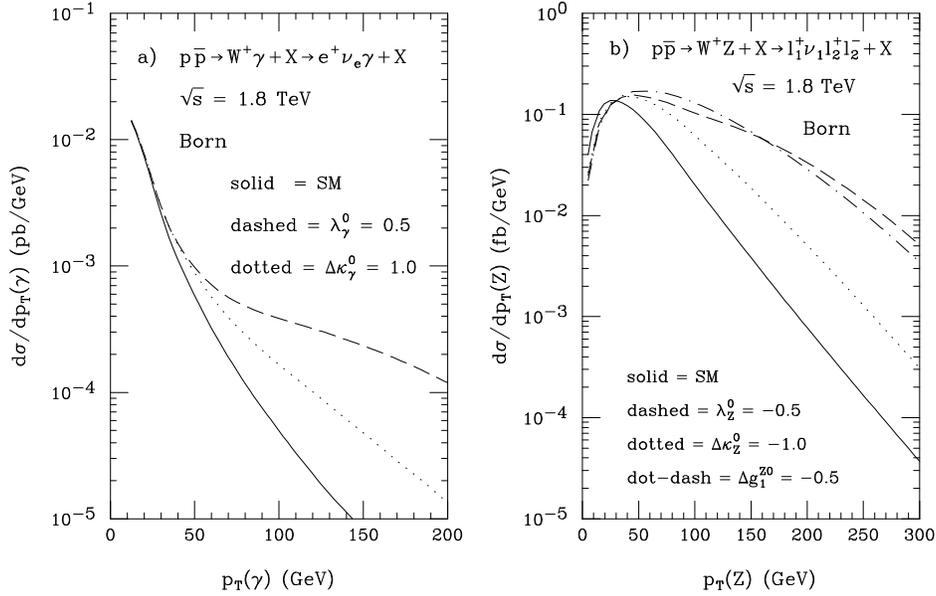}}
\rotl1
\vspace*{-0.35in}
\caption{Transverse momentum distribution of a) the photon in
$W^+\gamma$ production and b) the $Z$ in $W^+Z$ production in $p\bar p$
collisions at the Tevatron for various sets of anomalous coupling parameters.
{}From Ref.~\protect\cite{dpf}. }
\vspace*{-0.1in}
\end{center}
\end{figure}
Due to the more difficult background situation and also because of
insufficient knowledge of QCD radiative corrections~\cite{Ohnemus},
structure function
effects etc., a comparison of measured and theoretically predicted cross
sections at the ${\cal O}(10^{-2})$ level is not feasible at hadron colliders.
Rather their sensitivity to anomalous TGV's derives from fairly large
deviations from the SM in at least some regions of phase space. A typical
example is shown in Fig.~4 
where for most choices of anomalous
couplings $d\sigma/dp_T(V)$ is increased by one order of magnitude or more
in parts of the accessible transverse momentum range.

The actual
shape of the $p_T$ distributions depends crucially on the energy dependence
of the form factors. In weakly interacting models of new physics (like
supersymmetry~\cite{SUSY}) one should expect that virtual effects of heavy
particles (of mass $M$) will never lead to changes of cross sections by such
large factors. At low energy anomalous couplings are expected to scale
like~\cite{einhorn}
\begin{equation}
\Delta f_i(0) \sim {g^2\over 16\pi^2} {m_W^2\over M^2}\; ,
\end{equation}
while, at energies above the heavy particle mass $M$, form factor damping will
set in and qualitatively a behaviour like
\begin{equation}
\Delta f_i(s>>M^2) \sim {g^2\over 16\pi^2} {m_W^2\over s}
\end{equation}
must be expected. Turning to the effect on the weak boson pair production cross
section one finds that even when including a $\gamma^2=s/4m_W^2$ enhancement
factor in the amplitude, the amplitude changes only by a term of order
\begin{eqnarray}
\Delta{\cal M} \sim {s\over m_W^2} \Delta f_i(s) & \sim &
{g^2\over 16\pi^2} {s\over M^2}
\quad {\rm for}\;\; s<<M^2\;, \nonumber \\
&&{g^2\over 16\pi^2} {\phantom {s\over M^2} }\quad {\rm for}\;\; s>>M^2\;,
\end{eqnarray}
and the change in the amplitude is at most of the order of the naive
perturbative expectation, $g^2/16\pi^2$. Thus, for weakly coupled new
physics, no large enhancement of weak boson pair production cross sections
occurs. The dramatic increase of $WZ$ or $W\gamma$ production rates
shown in Fig.~4 
needs some strong interaction dynamics in the weak boson sector, and even
then it is not guaranteed to occur~\cite{einhorn}.

One thus finds that in their search for anomalous TGV's $e^+e^-$ and hadron
colliders are complementary. LEP or a 500~GeV NLC will probe $W^+W^-$ pair
production cross sections quite precisely, but at relatively low center of
mass energies. This limits the enhancement factors for anomalous TGV's but,
with sufficient statistics, even relatively weakly coupled new physics may
be accessible. At the hadron colliders much higher energy ranges can be
probed, but these experiments are only sensitive to strongly coupled new
physics.

\section*{4. Finite Width Effects and Gauge Invariance}

Some of the features of anomalous couplings, namely form factors and the
necessity to consider the full $S$-matrix elements can nicely be illustrated
by some very non-anomalous physics, namely fermion loop corrections within
the SM. At the same time I would like to address the problem of how to
implement finite width effects while maintaining gauge invariance when
dealing with processes involving TGV's~\cite{vO}. The discussion will
closely follow Ref.~\cite{BZwidth}.

Let us consider $W\gamma$ production at hadron colliders as an example.
Denoting the photon polarization vector by $\varepsilon^{*\mu}$ we can write
the amplitude as
\begin{equation}
{\cal M} = \varepsilon_\mu^* {\cal M}^\mu  =
\varepsilon_\mu^* {\cal M}_q^\mu +
\varepsilon_\mu^* {1\over \hat s -m_W^2} {\cal M}_W^\mu \; ,
\end{equation}
where ${\cal M}_q$ denotes $t$- and $u$-channel quark exchange graphs
and ${\cal M}_W$ stands for the $s$-channel $W$ exchange graph which
involves the TGV. Electromagnetic gauge invariance is guaranteed by
the relation
\begin{equation}
k_\mu {\cal M}^\mu = k_\mu {\cal M}_q^\mu +
k_\mu {1\over \hat s -m_W^2} {\cal M}_W^\mu = 0\; .
\end{equation}
Replacing the $W$ propagator factor by a Breit-Wigner form,
$1/(\hat s-m_W^2+im_W\Gamma_W)$,
disturbs the gauge cancellations between the individual Feynman graphs
and thus leads to an amplitude which is not electromagnetically gauge
invariant. In addition, a constant imaginary part in the inverse propagator
is ad hoc: it results from fermion loop contributions to the $W$ vacuum
polarization and the imaginary part should vanish for space-like momentum
transfers.

The general structure is best understood by first considering the lower
order process $q\bar q' \to W^-\to \ell^-\bar\nu$ without photon emission
(see Fig.~5). 
Finite width effects are included by resumming
the imaginary parts of the fermion loops. Neglecting fermion masses, the
transverse part of the $W$ vacuum polarization receives an imaginary
contribution
\begin{equation}\label{Pi_W^T}
Im\, \Pi_W^T(q^2) = \sum_f {g^2\over 48\pi} q^2
= q^2 {\Gamma_W\over m_W}\; ,
\end{equation}
while the imaginary part of the longitudinal piece vanishes. In the
unitary gauge and for $q^2>0$ the $W$ propagator is thus given by
\begin{eqnarray}
D_W^{\mu\nu}(q) & = &{-i \over q^2-m_W^2 + i Im\, \Pi_W^T(q^2)}
\left( g^{\mu\nu}-{q^\mu q^\nu \over q^2}\right)
+{i\over m_W^2 - i Im\, \Pi_W^L(q^2)}\; {q^\mu q^\nu \over q^2}
\nonumber \\[3.mm]
& = & {-i \over q^2-m_W^2 + i q^2 \gamma_W}
\left( g^{\mu\nu}-{q^\mu q^\nu \over m_W^2}
(1+ i \gamma_W )   \right)\; ,  \label{Wprop}
\end{eqnarray}
where the abbreviation $\gamma_W = \Gamma_W/m_W$ has been used. Note that
the $W$ propagator has received a $q^2$ dependent effective width which
actually would vanish in the space-like region.

\begin{figure}[h]
\epsfxsize=4.4in
\label{figwtree}
\begin{center}
\hspace*{0in}
\epsffile{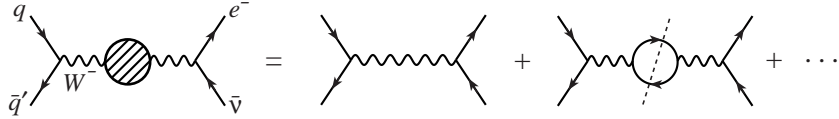}
\vspace*{0.15in}
\caption{Feynman graphs for the process $q\bar q'\to \ell^-\bar\nu$ at
lowest order. The resummation of the imaginary part of the $W$ vacuum
polarization leads to the Breit-Wigner type $W$ propagator of
Eq.~\protect\ref{Wprop} which is represented by the shaded blob.}
\vspace*{-0.15in}
\end{center}
\end{figure}

Now consider the same process, but including photon emission. A gauge
invariant expression is obtained by attaching the final state photon
in all possible ways to all charged particle propagators in the Feynman
graphs of Fig.~5. 
This includes radiation off the two incoming quark lines, radiation off
the final state charged lepton, and radiation off the $W$ propagators.
In addition, the photon must be attached to the charged fermions inside
the $W$ vacuum polarization loops, leading to the fermion triangle graphs
of Fig.~6. 
For a consistent treatment we only need to include the imaginary part
of the triangle graphs which is obtained by cutting the triangle graphs
into on-shell intermediate states in all possible ways, as shown in the
figure.

\begin{figure}
\epsfxsize=4.0in
\label{figvertex}
\begin{center}
\hspace*{0in}
\epsffile{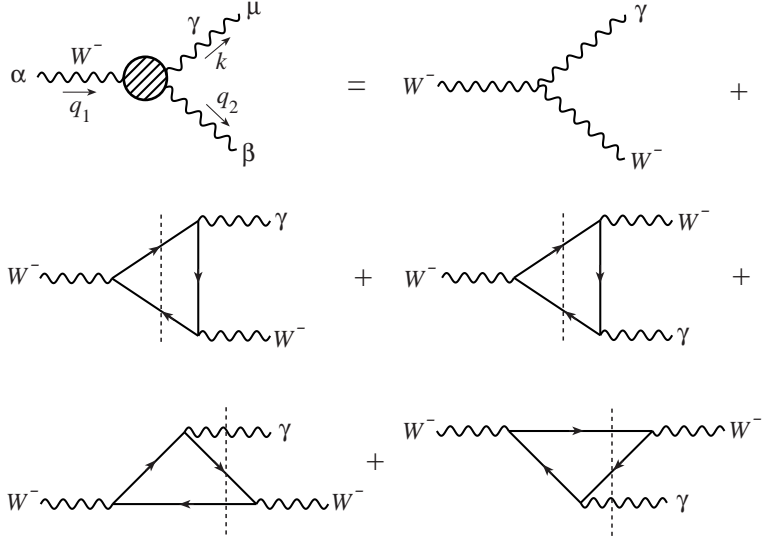}
\vspace*{0.15in}
\caption{
Effective $WW\gamma$ vertex as needed in the tree level calculation of
$W\gamma$ production. In addition to the lowest order vertex the imaginary
parts of the fermion triangles must be included
(see Eq.~\protect\ref{WWgvertex}).
}
\vspace*{-0.1in}
\end{center}
\end{figure}

For the momentum flow of Fig.~6 
the lowest order vertex is given by the familiar expression
\begin{equation}\label{LOWWgvertex}
-ie\Gamma_0^{\alpha\beta\mu} = -i e \left( (q_1+q_2)^\mu g^{\alpha\beta}
                            - (q_1+k)^\beta g^{\mu\alpha}
                            + (k-q_2)^\alpha g^{\mu\beta} \right)\; .
\end{equation}
Neglecting the masses of the fermions in the triangle graphs and dropping
terms proportional to $k^\mu$ (which will be contracted with the photon
polarization vector $\varepsilon^{*\mu}$ and hence vanish in the
amplitude) the contributions from the four triangle graphs reduce
to an extremely simple form. Each fermion doublet $f$, irrespective
of its hypercharge, adds $i(g^2/48\pi)\Gamma_0$ to the lowest order
$WW\gamma$ vertex $\Gamma_0$. After summing over all fermion species,
the lowest order vertex is thus replaced by
\begin{equation}\label{WWgvertex}
\Gamma^{\alpha\beta\mu} =
\Gamma_0^{\alpha\beta\mu}\left(1+\sum_f {ig^2\over 48 \pi}\right) =
\Gamma_0^{\alpha\beta\mu}\left(1+i\,{\Gamma_W\over m_W}\right) =
\Gamma_0^{\alpha\beta\mu}\left(1+i\gamma_W \right)\; .
\end{equation}

By construction, the resulting amplitude for the process
$q\bar q' \to \ell^- \bar\nu  \gamma $ is gauge invariant.
Indeed, gauge invariance of the full amplitude can be traced to the
electromagnetic Ward identity~\cite{lopez}
\begin{equation}\label{ward}
k_\mu {\Gamma_{\alpha\beta}}^\mu = \left(
iD_W\right)^{-1}_{\alpha\beta}(q_1)
- \left( iD_W\right)^{-1}_{\alpha\beta}(q_2)\; .
\end{equation}
Since
\begin{equation}\label{ward1}
k_\mu \Gamma^{\alpha\beta\mu} = \left(
( q_1^2 g^{\alpha\beta} - q_1^\alpha q_1^\beta ) -
( q_2^2 g^{\alpha\beta} - q_2^\alpha q_2^\beta ) \right)
\left(1+i\gamma_W\right)\;,
\end{equation}
and
\begin{equation}\label{ward2}
\left( iD_W\right)^{-1}_{\alpha\beta}(q) =
\left( q^2-m_W^2+iq^2\gamma_W \right)
\left( g_{\alpha\beta} - {q_\alpha q_\beta \over q^2} \right)
-m_W^2 {q_\alpha q_\beta \over q^2}\; ,
\end{equation}
this Ward identity is satisfied by our $W$ propagator and $WW\gamma$
vertex.

The modification of the lowest order $WW\gamma$ vertex in Eq.~\ref{WWgvertex}
looks like the introduction of anomalous couplings $g_1^\gamma = \kappa_\gamma
= 1+i\gamma_W$ and one may thus worry that the full amplitude will violate
unitarity at large center of mass energies $\sqrt{\hat s}$. While indeed the
vertex is modified, this modification is compensated by the effective
$\hat s$-dependent width in the propagator. As compared to the expressions
with a lowest order propagator, $1/(\hat s - m_W^2)$,  which of course has
good high energy behaviour, the overall effect is multiplication of the
$s$-channel $W$-exchange amplitude ${\cal M}_W$ by a factor
\begin{equation}
G(\hat s) = {\hat s - m_W^2 \over \hat s(1+i\gamma_W) - m_W^2}\;
(1+i\gamma_W) = 1\; -\;
{i\Gamma_W m_W \over \hat s - m_W^2 + im_W\Gamma_W{\hat s\over m_W^2}}\; .
\label{formfactor}
\end{equation}
Obviously, $G(\hat s)\to 1$ as $\hat s \to \infty$ and the high energy
behaviour of our finite width amplitude is identical to the one of the
naive tree level result for $W\gamma$ production. In fact, the contributions
from the triangle graphs are crucial to compensate the bad high energy
behaviour introduced by the $q^2$-dependent width in Eq.~\ref{Wprop}.

This interplay of propagator and vertex corrections illustrates the remarks
made in Section~2. The leading one-loop contributions, namely the imaginary
parts of $WW\gamma$ vertex and inverse $W$ propagator, lead to a change of
the $S$-matrix element for $W\gamma$ production which can be parameterized
in terms of the generalized vertex function
$\Gamma^{\mu\alpha\beta}_{\gamma WW}(k,q_1,q_2)$. The nonvanishing form
factors in its tensor decomposition are given by
\begin{equation}
g_1^\gamma(q_1^2)=\kappa_\gamma(q_1^2) = G(q_1^2) = 1\; -\;
{i\Gamma_W m_W \over q_1^2 - m_W^2 + i{\Gamma_W\over m_W} q_1^2}
\; , \label{ffexplicit}
\end{equation}
and the form factor scale is set by the masses of the particles
involved, here the $W$ boson mass.

\section*{5. Conclusions}

The direct measurement of the nonabelian $WW\gamma$ and $WWZ$ vertices
at present and future hadron and $e^+e^-$ colliders constitutes an
important test of the basic structure of electroweak interactions.
There are strong theoretical arguments that experiments will yield
exactly the results predicted by the SM, even though no rigorous
proof of this assertion exists. Observation of anomalous
couplings at either the Tevatron or at LEP II would therefore have
grave consequences for our understanding of electroweak physics.

Irrespective of how likely an observation of anomalous couplings
might be, $e^+e-$ and hadron colliders measure very different
aspects of $WWV$ vertex functions. With sufficient statistics $e^+e^-$
experiments are able to probe small deviations from SM cross sections
and are hence sensitive to weakly interacting new physics. However,
they will mainly probe just one process, $W$ pair production, in
a limited energy range. Hadron colliders, on the other hand, can
investigate all electroweak boson pair production processes, albeit
with lower accuracy. They look for relatively large enhancements of
cross sections at high center of mass energies and are thus only
sensitive to new strong interaction dynamics in the bosonic sector.
Hadron and $e^+e^-$ machines are indeed complementary means to
directly study the nonabelian aspects of electroweak interactions.

\section*{Acknowledgements}
This research was supported in part by the U.S.~Department of Energy under
Grant No.~DE-FG02-95ER40896 and in part by the University of Wisconsin Research
Committee with funds granted by the Wisconsin Alumni Research Foundation.

\end{document}